\documentclass[a4paper,11pt]{article}
\usepackage{jheppub}
\usepackage{defs}
\usepackage{orcidlink}
\preprint{MLQC4FC-2026-001}
\title{\boldmath Hyperoptimisation algorithm for the next generation of PDF determinations:
ensemble regression with an unbiased selection model}

\author[1]{Juan M. Cruz-Martinez~\!\orcidlink{0000-0002-8061-1965}}
\author[2]{Tommaso Giani~\!\orcidlink{0000-0001-7135-6731}}
\author[3]{Tanjona R. Rabemananjara~\!\orcidlink{0000-0002-8395-8059}}

\affiliation[1]{Departamento de Física Atómica, Molecular y Nuclear, Universidad de Sevilla, E-41080 Sevilla, Spain}
\affiliation[2]{Dipartimento di Fisica, Universit\`a degli Studi di Torino and INFN, Sezione di Torino, Via Pietro Giuria 1, I-10125 Torino, Italy}
\affiliation[3]{Université Paris-Saclay, CNRS, IJCLab, 91405 Orsay, France}

\emailAdd{jcruz@us.es}
\emailAdd{tommaso.giani@unito.it}
\emailAdd{tanjona.rabemananjara@ijclab.in2p3.fr}

\abstract{
We present a new automated procedure for selecting fitting methodologies in the determination of
parton distribution functions (PDFs), based on a hyperoptimisation algorithm using ensemble regression.
Building on the $k$-folding approach previously employed by the NNPDF collaboration, we introduce a
systematic strategy to generate ensembles of hyperparameter configurations and define a new $k$-folding
metric that consistently incorporates full PDF uncertainties.
The algorithm outputs an ensemble of statistically equivalent fitting methodologies, which are combined
to produce a single PDF set whose uncertainties consistently account for hyperparameter variation.
We assess the impact of this approach by comparing results obtained with identical data and theoretical
inputs but different fitting methodologies, determined using the previous and the new hyperoptimisation
procedures. The new method broadly confirms earlier NNPDF results, while yielding moderately larger
uncertainties in regions where data provide limited constraints.
}

\begin{document}
\maketitle
\flushbottom

\section{Introduction}

Parton distribution functions (PDFs) are essential inputs for theoretical predictions and experimental analyses
at hadron colliders, encoding the non-perturbative structure of the proton. Since PDFs cannot be computed from
first principles, they are determined through global analyses of experimental data within a given theoretical 
framework~\cite{Hou:2019efy, Bailey:2020ooq, NNPDF:2021njg}. As a result, they inherit uncertainties from both
data and theory, compounded by limited knowledge of their underlying functional form. Consequently, PDFs remain
among the most extensively studied yet least precisely known ingredients in precision high-energy physics. This
is reflected in recent analyses at the Large Hadron Collider (LHC), where PDF uncertainties constitute one of
the dominant sources of error~\cite{CMS:2024ony, ATLAS:2023fsi, CMS:2024lrd}.

Several methodologies have been developed for PDF determination. A widely used approach, adopted for instance by
the CT~\cite{Hou:2019efy} and MSHT~\cite{Bailey:2020ooq} collaborations, relies on fixed functional forms within
a Hessian framework. In this case, PDFs are parametrised using a limited set of parameters, and the best fit is 
obtained by minimising the experimental $\chi^2$. Uncertainties are then derived by expanding around the minimum 
using the Hessian covariance matrix, typically with an enlarged tolerance criterion $\Delta\chi^2 = T$ to account
for tensions in data and theoretical limitations, effectively increasing the size of the PDF uncertainties.
An alternative methodology, developed by the NNPDF collaboration~\cite{NNPDF:2021njg}, employs neural networks as 
flexible, unbiased function approximants. In this framework, uncertainties are estimated using a Monte Carlo replica 
method where ensembles of PDF replicas are obtained by fitting neural networks to corresponding replicas of the 
experimental data, which encode all information about theoretical and experimental uncertainties as well as 
correlations. Each PDF replica is fitted to a different data replica, through the minimization of a suitable loss
and the implementation of a training/validation split procedure, as customary in machine learning fitting algorithms. 
The final output of the analysis is a set of trained PDF replicas, which can be used to compute central value and 
uncertainty of any observable function of the PDF.

While neural networks reduce parametrisation bias, the methodology remains dependent on the choice of hyperparameters, i.e., a set of methodological settings such as the neural network architecture and the optimizer specifications, which define the space of functions accessible to the model. As a result, manual selection introduces arbitrariness 
and potential biases. To address this shortcoming, NNPDF has employed an automated hyperparameter 
optimisation~\cite{Carrazza_2019} in which the choice of hyperparameters is dictated by the ability of the network
to describe well the training data while keeping its generalisation power for unseen data. This reduces both 
underfitting and overfitting and eliminates the human selection bias.
However, due to the substantial computational cost of training thousands of models, the hyperoptimisation strategy of 
the latest NNPDF release was subject to several simplifications. First, the search for the best hyperparameters was 
limited to the central value PDF, rather than looking at the full ensembles of trained replicas. This meant that the 
optimisation algorithm had no access to the uncertainties of the PDF itself, so the generalisation power could only be 
evaluated on the experimental uncertainties. Second, once a best set of hyperparameters was selected, this was used for 
all replicas in the final PDF set and for all associated studies.  Therefore, while this approach represented a 
significant step forward, the determination neglects the PDF error during the optimisation of the hyperparameters and, 
crucially, does not account for the uncertainty associated with the choice of the hyperparameters themselves in the 
final PDF set. In Ref.~\cite{NNPDF:2021njg}, the impact of these limitations was assessed through a series of
\emph{a posteriori} consistency checks.

Recent advances in computational resources and methodology~\cite{Cruz-Martinez:2024wiu} have made it possible to 
overcome these limitations. Ref.~\cite{Cruz-Martinez:2024wiu} presented a preliminary study demonstrating the access
to the entire ensemble of PDF replicas by the hyperoptimisation algorithm, as well as the notion of sampling over 
different hyperparameter choices.
In this work, we extend~\cite{Cruz-Martinez:2024wiu} by introducing a new log-likelihood-based figure of merit that 
fully incorporates the PDF ensemble without favouring artificially large uncertainties. Additionally, we also develop
a systematic procedure to sample and combine multiple statistically equivalent hyperparameter configurations, whose
combination is employed to construct the final PDF distribution. This allows the final PDF uncertainties to 
consistently include the contribution from hyperparameter variation, thereby reducing residual methodological bias.
The new algorithm is implemented in the public NNPDF code, integrated with the full fitting framework, and will be
used for future PDF determinations by the NNPDF collaboration. The important computational burden introduced by the
new procedure is handled thanks to the access to clusters of GPUs.

The paper is structured as follows. In Section~\ref{sec:nnpdf_methodology} we briefly review the necessary concepts
underlying the NNPDF methodology in order to make this paper self-contained. Section~\ref{sec:hyperscan} introduces
the new hyperoptimisation algorithm and its underlying figure of merit, highlighting its conceptual advances over 
Refs.~\cite{NNPDF:2021njg, Cruz-Martinez:2024wiu}. In Section~\ref{sec:results} we present a phenomenological study 
exploring the interplay between hyperparameter optimisation and theoretical inputs. Conclusions are given in 
Section~\ref{sec:conclusions}.

\section{Summary of the NNPDF methodology}
\label{sec:nnpdf_methodology}

In this Section we recall the main features of the NNPDF fitting methodology. We discuss the usage of neural networks 
as representation for PDF probability distributions, the general training procedure, and the NNPDF approach to generate 
samples from the trained networks.

\paragraph{Prior distributions for PDFs.} PDFs are defined within the space of continuous functions having support in 
the interval $\left[0,1\right]$, and satisfying a number of theoretical constraints, such as integrability conditions 
and sum rules. Despite this being an infinite dimensional space, it is possible to build a probability measure on it, 
and introduce a prior probability distribution for the PDFs. Following the discussion of Ref.~\cite{DelDebbio:2021whr}, 
one could recast the problem in finite dimensions by considering the values of the PDF $f$ at selected values of
$\mathbf{x}$.
\begin{align}
    \label{eq:xgrid}
    \mathbf{x}=\{x_i; i=1,
    \ldots, N\}\,,
\end{align}
defining a N-dimensional vector of stochastic variables
\begin{align}
    \label{eq:FunVect}
    \mathbf{f} = f(\mathbf{x}) = 
    \begin{pmatrix}
        f(x_1) \\
        \vdots \\
        f(x_N)
    \end{pmatrix} \in \mathbb{R}^N\,.
\end{align}
In the NNPDF approach, the probability distribution of $\mathbf{f}$ is represented by means of a neural network
$\text{NN}\left(x;\mathbf{w}\right)$, defined on the domain $x\in \left[0,1\right]$ and built from a large number
($\sim 800$) of weights $\mathbf{w}$.
In practice, different samples for the variable $\mathbf{f}$ can be generated by random initialisation of the weights, 
and considering the network values on the grid $\mathbf{x}$. In this sense, the neural network effectively represents a 
probability distribution for the variable $\mathbf{f} \in\mathbb{R}^N$. We denote such probability distribution as
$\text{NN}\left(\mathbf{f}\right)$.
An alternative approach to define a probability distribution for PDF is presented in Ref.~\cite{Candido:2024hjt} where, 
rather than using a neural network, the probability distribution of $\mathbf{f}$ is given analytically using the 
formalism of Gaussian Processes.

\paragraph{Monte Carlo replica method.} 
An ensemble of PDF samples 
\begin{align}
    \{\mathbf{f}^{(k)}\,, k=1,...,N_{rep}\}\,,
\end{align}
are generated from the distribution $\text{NN}\left(\mathbf{f}\right)$ by randomly initializing the network weights $\mathbf{w}$.
Denoting as $y$ the central values of the data entering the global analysis, an equal number $N_{rep}$ of experimental data replicas $y^{(k)}$
\begin{align}
    \{y^{(k)}\,, k=1,...,N_{rep}\}\,,
\end{align}
are drawn from the experimental distributions.
Each PDF replica $\mathbf{f}^{(k)}$
is then fitted to a different data replica $y^{(k)}$.
This is done through a gradient descent algorithm or, before Ref.~\cite{NNPDF:2021njg}, a genetic optimizer.
We also implement a training/validation split procedure: the experimental data are split into two disjoint sets, which are used to build a training and validation loss. The former is minimized as a function of the network weights $\mathbf{w}$, until the point at which the validation loss stops decreasing.
In practice, this translates into $N_{rep}$ (usually $\sim 100$) independent optimisation problems on the weights $\mathbf{w}$ of the neural network, and delivers as output an ensemble of $N_{rep}$ trained PDF replicas $\mathbf{f^*}^{(k)}$.
These can be seen as independent samples from the final distribution of PDF replicas conditioned to the observed experimental values. We denote such distribution as $p\left(\mathbf{f}|y\right)$.
The PDF central value and uncertainty can be computed as
\begin{align}
\label{eq:mean_cov_pdf}
\text{E}\left[\mathbf{f}^*\right] = \frac{1}{N_{rep}}\sum_k \mathbf{f^*}^{(k)}\,,\quad
    \text{cov}[\mathbf{f}^*] = \frac{1}{N_{rep}}\sum_k \left(\mathbf{f^*}^{(k)}-\text{E}\left[\mathbf{f}^*\right]\right)
    \left(\mathbf{f^*}^{(k)}-\text{E}\left[\mathbf{f}^*\right]\right)^T\,,
\end{align}
alongside central values and uncertainties for any observable function of the PDFs.
As stressed in Ref.~\cite{DelDebbio:2021whr,Costantini:2024wby},
the Monte Carlo replicas approach correctly propagates experimental uncertainties only when the observables entering the fit are linear in the PDF.
Nevertheless, also in the case of non-linear models, the MC approach is expected to work at least in proximity of $\text{E}\left[\mathbf{f}^*\right]$, around which a linear approximation of the physical observables can be done.
This quasi-linearity is also assumed in the hessian approach~\cite{Watt:2012tq}, and indeed both methods produce consistent results~\cite{Watt:2012tq,Harland-Lang:2024kvt}.
We refer the reader to Section~3.2 of Ref.~\cite{DelDebbio:2021whr} for a complete discussion of this issue. 

\section{Hyperparameters selection}
\label{sec:hyperscan}

The fitting methodology described above is built according to a specific choice of hyperparameters.
Some of these define the architecture of the network (number of nodes, of layers, activation functions, etc.),
and other the optimisation (minimiser algorithm, learning rate, etc.).
In the case of a PDF fit, we include as hyperparameters also theoretical settings such as the strength by which certain physical constraints are imposed, as they can alter the optimisation landscape.
Such hyperparameters, denoted as $\mathbf{\theta}$,  effectively control different features of the initial PDF distribution $\text{NN}\left(\mathbf{f}\right)$, and the way in which the training takes place. 
The final distribution of trained replicas $p\left(\mathbf{f}| y\right)$ will therefore depend on $\theta$, and should be seen as a probability distribution conditioned on certain methodological settings, $p\left(\mathbf{f}|y,\theta\right)$. 
It is therefore important to develop a systematic way to include effects due to hyperparameter variations in the final results.
In order to address this point, let us assume that the hyperparameters $\mathbf{\theta}$ are distributed according to some probability distribution $\eta\left(\theta\right)$. The central value and uncertainty of the PDF can be computed as the mean and variance of the variable $\mathbf{f}$:
\begin{align}
    \label{eq:mc_mean}
    \langle \mbf \rangle_{\mbf,\theta} &= 
    \int d\theta \left[\int d \mbf \, \mbf \, p\left( \mbf| \theta, y \right) \right]\eta\left( \theta\right)
    = \int d\theta \, \mathbf{ m}\left(\theta\right) \eta\left( \theta\right)\,,\\
    \nonumber
    \langle \left(\mbf - \langle \mbf \rangle_{\mbf,\theta}\right)^2 \rangle_{\mbf,\theta}   
      &= \int d\theta \, \eta\left(\theta \right) \int d \mbf  \,  p\left( \mbf| y, \theta \right) \left(\mbf - \langle \mathbf{ m}\left(\theta\right)\rangle_{\theta}\right)^2 \\
      \label{eq:mc_error_decomposed}
      &=\int d\theta \, \eta\left(\theta \right) 
\left(K\left(\theta \right) + \left(\mathbf{ m}\left(\theta\right) -\langle \mathbf{ m}\left(\theta\right) \rangle_{\theta} \right)^2\right) \,,
\end{align}
where $\mathbf{m}\left(\theta\right)$ denotes the central value of the PDF for fixed values of $\theta$.
In order to get Eq.~\eqref{eq:mc_error_decomposed}, we have assumed the trained replica distribution $p\left( \mbf| y, \theta \right)$ to be a multigaussian having mean $\mathbf{ m}\left(\theta\right)$ and covariance $K\left(\theta\right)$.
Eqs.~\eqref{eq:mc_mean} and~\eqref{eq:mc_error_decomposed} show how mean and covariance of the PDF are given by integrals on the hyperparameters distribution $\eta\left(\theta\right)$.
As discussed in Ref.~\cite{Medrano:2025cmg},  Eq.~\eqref{eq:mc_error_decomposed} allows to visualize the PDF error as the sum (in quadrature) of two components, the first given by the covariance $K\left(\theta\right)$, representing the PDF uncertainty for fixed value of $\theta$, the second taking into account the variance of the PDF central value when varying $\theta$.
By fixing the hyperparameters to a certain value, this second contribution is neglected.

\subsection{Hyperoptimisation metrics}
\label{sec:hyperopt_metrics}
\paragraph{$k$-folding.} Hyperparameters selection is designed in order to maximize the generalisation power of the methodology. This is achieved by implementing $k$-folding: the full dataset is split in $k$ disjoint subsets, denoted as folds. Folds are constructed in such a way that the data contained in each of them are representative of the full global dataset, both in terms of physical processes and kinematical coverage. For each chosen point $\theta$ in hyperparameter space, $k$ fits are performed, each time excluding one of the folds and using the remaining $(k-1)$ ones to train the network. The data entering the fit are then split in training and validation sets, as customary in the NNPDF methodology. 
A specific metric is defined in order to quantify the performance of the resulting fits on the excluded folds. This is taken as a measure of the generalisation power of the methodology.
Hyperparameters can then be selected according to the performance of each methodology in unseen (or seen) data.

\paragraph{NNPDF4.0 hyperoptimisation metric.} In NNPDF4.0~\cite{NNPDF:2021njg}, the hyperoptimisation metric was defined starting from the $\chi^2$ of the excluded folds: denoting as $y^{(i)}\,, C^{(i)}_y$ the central values of the fold $i$ and its corresponding experimental covariance matrix, and as $\mbf_{\{-i\}}$ a fit from which the fold $i$ has been excluded, the $\chi^2$ of the excluded fold is given by
\begin{align}
    \label{eq:chi2_excluded_fold}
    \chi_i^2\left(\theta\right) = \left(y^{(i)} - T^{(i)}\left(\theta\right)\right)
    \left[C^{(i)}_{y}\right]^{-1}\left(y^{(i)} - T^{(i)}\left(\theta\right)\right)^T\,,
\end{align}
where $T^{(i)}\left(\theta\right)$ denotes the theory prediction for $y^{(i)}$, computed using the fit $\mbf_{\{-i\}}$.
The hyperoptimisation metric in NNPDF4.0 was defined as the average of the $\chi^2_i$ across all folds,
\begin{align}
    \label{eq:nnpdf4.0_hyperopt_metric}
    \mathcal{L}_{\text{NNPDF4.0}}\left(\theta\right) 
    = \frac{1}{k} \sum_{i}^k \chi^2_i\left(\theta\right)\,.
\end{align}
Due to computing limitations, $T^{(i)}\left(\theta\right)$ was not computed as the central value of an ensemble of fits as per Eq.~\eqref{eq:mean_cov_pdf}, but rather starting from a single fit to central data for each fold.
Eq.~\eqref{eq:nnpdf4.0_hyperopt_metric} was then minimized in order to identify one single best configuration of hyperparameters.
Note how Eq.~\eqref{eq:nnpdf4.0_hyperopt_metric} does not correspond to the full log-likelihood of the excluded fold. As we will discuss in the next paragraph, this would require the knowledge of the PDF error for which a whole ensemble of replicas have to be run for each datapoint $\theta$ in hyperparameter space.
This task has been made computationally feasible only recently.

\paragraph{New hyperoptimisation metric for future PDF releases.} 
As mentioned in the introduction, thanks to technical improvements and the access to clusters of GPUs, the computational limitations affecting NNPDF4.0 hyperoptimisation have now been lifted. 
For each point in hyperparameter space a full ensemble of Monte Carlo replicas can now be produced, giving access to the full PDF distribution also during the hyperparameter optimisation procedure.
Theory predictions for the point $y^{(i)}$ can now be computed as an expectation value over the replicas 
\begin{align}
    \overline{T}^{(i)}\left(\theta\right) =  <T^{(i)}\left(\theta\right)>_{\text{replicas}}\,,
\end{align}
and the PDF error can be accessed by computing the covariance over replicas $C^{(i)}_{\text{PDF}}\left(\theta\right)$, given by
\begin{align}                       
\left(C^{(i)}_{\text{PDF}}\left(\theta\right)\right)_{\alpha \beta} = \cov\left[T^{(i)}_{\alpha}\left(\theta\right), T^{(i)}_{\beta}\left(\theta\right)\right]\,.
\end{align}
Having this information the full likelihood can now be computed.
Assuming that both experimental data and theory predictions are Gaussian-distributed, the probability of the fold $i$ given the fit $\mbf_{\{-i\}}$ and the hyperparameters $\theta$ is given by
\begin{align}
    \label{eq:likelihood}
    p\left(y^{(i)} | \mbf_{\{-i\}}, \theta\right) = \frac{1}{N\left(\theta\right)}
    e^{-\frac{1}{2}\left(y^{(i)} - \overline{T}^{(i)}\left(\theta\right)\right)
    \left(C^{(i)}_y + C^{(i)}_{\text{PDF}}\left(\theta\right)\right)^{-1}\left(y^{(i)} - \overline{T}^{(i)}\left(\theta\right)\right)^T}\,.
\end{align}
The term $N\left(\theta\right)$ is a normalization factor necessary for the likelihood to integrate to 1, and it is therefore given by
\begin{align}
    \label{eq:likelihood_normalization}
    N\left(\theta\right) =
    \sqrt{\det{2\pi\left(C^{(i)}_y + C^{(i)}_{\text{PDF}}\left(\theta\right)\right)}}\,.
\end{align}
Crucially, as a consequence of the presence of the PDF error $C^{(i)}_{\text{PDF}}\left(\theta\right)$, this normalization depends on the hyperparameters $\theta$. 
The loss defined in Eq.~\eqref{eq:chi2_excluded_fold}
can now be substituted with the full log-likelihood of the excluded fold,
\begin{align}
    \mathcal{L}^{(i)}\left(\theta\right) &\equiv - 2 \, \log p\left(y^{(i)} | \mbf_{\{-i\}}, \theta\right) \nonumber \\
    &= \left(y^{(i)} - \overline{T}^{(i)}\left(\theta\right)\right) 
    \left(C^{(i)}_y + C^{(i)}_{\text{PDF}}\left(\theta\right)\right)^{-1}\left(y^{(i)} - \overline{T}^{(i)}\left(\theta\right)\right)^T  \label{eq:log_likelihood} \\ 
    &\quad \quad+ \log \det \left(C^{(i)}_y + C^{(i)}_{\text{PDF}}\left(\theta\right)\right)\,. \nonumber
\end{align}
This log-likelihood is given by the sum of two terms.
The first can be readily identified with Eq.~\eqref{eq:chi2_excluded_fold}, but now including the PDF uncertainties,
\begin{align}
    \label{eq:chi2term}
    \chi_{(i)}^2\left(\theta\right) = \left(y^{(i)} - \overline{T}^{(i)}\left(\theta\right)\right) 
    \left(C^{(i)}_y + C^{(i)}_{\text{PDF}}\left(\theta\right)\right)^{-1}\left(y^{(i)} - \overline{T}^{(i)}\left(\theta\right)\right)^T\,,
\end{align}
which is the figure of merit used in Ref.~\cite{Cruz-Martinez:2024wiu}.
The second term instead arises from the normalization of the likelihood $N\left(\theta\right)$ and it is often referred to in the literature as the complexity penalty,
\begin{align}
    \label{eq:penalty}
    \mathcal{P}^{(i)}\left(\theta\right) = \log \det \left(C^{(i)}_y + C^{(i)}_{\text{PDF}}\left(\theta\right)\right)\,.
\end{align}
Qualitatively speaking, Eq.~\eqref{eq:penalty} can be seen as a penalty for models achieving a good loss thanks to an unnaturally large error: small values for the loss of Eq.~\eqref{eq:chi2term} could also be achieved thanks to a large PDF error, and not because a good description of the data is provided by the theory predictions $\overline{T}^{(i)}\left(\theta\right)$.
How the complexity penalty prevents such a situation to happen can be clearly seen by moving to the basis of the eigenvectors of the total covariance matrix.
In this case, dropping the fold index $i$ and denoting as $v\left(\theta\right)$  and $\lambda_{\alpha}\left(\theta\right)$ its eigenvectors and eigenvalues respectively we have
\begin{align}
    \label{eq:diagonal_case}
    \chi^{2}\left(\theta\right) = \sum_{\alpha}\frac{\left(v_{\alpha}\left(\theta\right)\right)^2}{\lambda_{\alpha}\left(\theta\right)}\,,\quad\quad
    \mathcal{P}\left(\theta\right) = \log \prod_{\alpha} \lambda_{\alpha}\left(\theta\right)\,,
\end{align}
from which it is evident how a large PDF error increasing $\lambda_{\alpha}\left(\theta\right)$ (and reducing therefore $\chi^2\left(\theta\right)$) would correspond to a larger penalty $\mathcal{P}\left(\theta\right)$. 

The complexity penalty term arises naturally from the normalization of the likelihood of Eq.~\eqref{eq:likelihood_normalization}: since this depends on $\theta$ through the PDF error, when looking at the likelihood as a function of the hyperparameter, it is crucial to include it.
In the NNPDF4.0 metric, Eq.~\eqref{eq:nnpdf4.0_hyperopt_metric}, the PDF error did not enter the definition and therefore the likelihood normalization didn't have any dependence on $\theta$.
In Eq.~(2.19) of Ref.~\cite{Cruz-Martinez:2024wiu} instead it was not included despite the dependence in the definition of the $\chi^{2}$, introducing an implicit bias in the hyperoptimisation algorithm towards methodologies that increase the PDF uncertainties.

The practical implementation of Eq.~\eqref{eq:penalty} 
is performed in terms of dimensionless quantities only, by rescaling the total covariance matrix entering the log determinant by the experimental central values
\begin{align}
    \left(C^{(i)}_y + C^{(i)}_{\text{PDF}} \right)_{\alpha \beta} \longrightarrow \frac{\left(C^{(i)}_y + C^{(i)}_{\text{PDF}} \right)_{\alpha \beta}}{y^{(i)}_{\alpha}y^{(i)}_{\beta}}\,. \nonumber
\end{align}
In this way the numerical value of the log determinant will not be driven by the size of the absolute error, which can change a lot depending on the dataset.
We refer the reader to Chapter.~5 of Ref.~\cite{Rasmussen2006Gaussian} for a complete and more general discussion of the complexity penalty term.

The final hyperoptimisation metric is defined as the average over the log-likelihood of all excluded folds:
\begin{align}
    \label{eq:nnpdf4.1_hyperopt_metric}
    \mathcal{L}\left(\theta\right) = \frac{1}{k}\sum_{i}^k\mathcal{L}^{(i)}\left(\theta\right)\,,
\end{align}
analogously to $\mathcal{L}_{\text{NNPDF4.0}}$ Eq.~\eqref{eq:nnpdf4.0_hyperopt_metric}.

We conclude this Section by stressing that, in order for the algorithm to work as expected, fits performed during $k$-folding, denoted as $\mathbf{f}_{\{-i\}}$ in Eq.~\eqref{eq:likelihood}, should converge.
In practice this can be checked by monitoring the corresponding experimental $\chi_{\text{exp}}^2$. In order to speed up the convergence, a penalty term penalizing trials for which $\chi_{\text{exp}}^2>2$ can be added to the metric of Eq.~\eqref{eq:nnpdf4.1_hyperopt_metric}.


\subsection{Sampling of hyperparameter configurations}
\label{sec:sampling}
As mentioned in Sec.~\ref{sec:hyperopt_metrics}, in the context of NNPDF4.0, Eq.~\eqref{eq:nnpdf4.0_hyperopt_metric} was minimized in order to identify one single best configuration
\begin{align}
    \label{eq:theta_best_40}
    \mathbf{\theta}^* = \text{arg}\,\,\min_{\mathbf{\theta}} \, \mathcal{L}_{\text{NNPDF4.0}}\left(\theta\right)\,.
\end{align}
In doing so, the effect on the final PDF uncertainty given by the second term of Eq.~\eqref{eq:mc_error_decomposed} was neglected.
In this work, starting from the new metric introduced in Sec.~\ref{sec:hyperopt_metrics} and from a given prior in $\theta$ space that for examples defines the domain of the scanned hyperparameters, we generate an ensemble of samples, each one representing a different methodology
\begin{align}
    \label{eq:trials}
    \{\theta_i\,, i = 1,...,N_{trials}\}\,,
\end{align}
which can then be used to compute the integrals over $d\theta$ of Eqs.~\eqref{eq:mc_mean},~\eqref{eq:mc_error_decomposed}.
Such sampling is practically achieved by generating each PDF replica ${\mathbf{f}^*}^{(h)}$ using a different trial $\mathbf{\theta}_i$:
PDF replicas acquire an additional index $i$ labeling the specific hyperparameter configuration, and can be seen as samples from the joint probability distribution $p\left(\mathbf{f},\theta|y\right)$.
The integral over $d\theta$ is implicitly included when averaging over different replicas. 

\paragraph{TPE algorithm and generation of trials.} 
As already mentioned at the beginning of this Section, the hyperparameters $\theta$ include discrete variables (e.g.\ number of layers, activation functions), 
continuous variables (learning rate, clipnorm), and conditional choices (e.g.\ optimizer type with 
optimizer-specific parameters). The resulting search space is therefore heterogeneous and tree-structured.
Its exploration is performed using the Tree-structured Parzen Estimator (TPE)
algorithm as implemented in the \texttt{hyperopt} library~\cite{pmlr-v28-bergstra13}. TPE is a sequential model--based optimisation strategy designed to efficiently minimize expensive, objective functions, such as the metric $\mathcal{L}$ defined in Eq.~\eqref{eq:nnpdf4.1_hyperopt_metric}. In our case, a single evaluation of the loss requires the full training of neural networks across folds and replicas, thus
making brute-force or grid-based scans computationally prohibitive.
TPE constructs a probabilistic surrogate model for the objective function and uses it to select the most promising next evaluation point in hyperparameter space:
by modelling the probability density $p(\theta|\mathcal{L})$, and accumulating evaluations of $(\theta_j, \mathcal{L}_j)$ at each iteration $j$, new hyperparameter configurations are sampled preferentially in regions dictated by $p(\theta|\mathcal{L})$.
The first iterations are sampled randomly to initialize the density estimators (“thermalisation phase”), but after this initial exploration the algorithm
enters an exploitation regime where progressively better regions of hyperparameter space are identified and refined (see Figs.~\ref{fig:logp_evol} and~\ref{fig:logp_flav} in the following sections). 
The final output of the algorithm is therefore a chain of possible hyperparameter settings $\{\theta_i\}$, denoted as \textit{trials}, whose distribution progressively peaks around 
\begin{equation}
    \theta^\star = \arg \min_\theta \mathcal{L}(\theta).
\end{equation}
The TPE algorithm was already used in the context of NNPDF4.0 hyperoptimisation~\cite{NNPDF:2021njg} alongside the metric of Eq.~\eqref{eq:nnpdf4.0_hyperopt_metric} but could not be run long enough for a thermalization phase due to the aforementioned computational limitations.
Instead, the selection of the (single) final trial was performed by parallel scans of thousands of configurations.
This was inefficient as it generated many outliers as well as no clear hierarchy of configurations, relying only on the large amount of models.
In this sense, the TPE algorithm was not exploited in its full potential, and was instead used as a random scan of hyperparameter space.
To prevent this issue in the study of Ref.~\cite{Cruz-Martinez:2024wiu}, where the chain of trials would have not been allowed to achieve thermalisation, a restricted hyperparameter space was used to avoid outliers.

Thanks to our current access to clusters of GPUs, and the implementation of distributed training in NNPDF, we can now run the algorithm long enough to fully take advantage of its exploitation regime, and we take the thermalised chain of trials as the candidate for the ensemble introduced in Eq.~\eqref{eq:trials}.
Further details concerning the TPE algorithm are reported in App.~\ref{app:TPE}

\section{Results}
\label{sec:results}
The new metric presented in Sec.~\ref{sec:hyperopt_metrics} has been implemented in the public NNPDF framework~\cite{nnpdf40code}, alongside the sampling procedure described in Sec.~\ref{sec:sampling}. 
The TPE algorithm is run on GPU clusters to produce a chain of $\sim 1000$ trials, by taking as input a prior for the hyperparameters, specified in an input runcard, and the metric to be used (in our case Eq.~\eqref{eq:nnpdf4.1_hyperopt_metric}). 
The implementation is modular, making it straightforward to consider different metrics and alternative minimisation libraries in place of \texttt{hyperopt}.
A subset of the trials produced after thermalisation is considered, and used during the fit to generate different PDF replicas, to practically implement Eqs.~\eqref{eq:mc_mean},~\eqref{eq:mc_error_decomposed}.
In this Section we present our results.
First we discuss the differences among the best trials selected by \texttt{hyperopt}: by producing single trial fits we study the impact of different hyperparameter choices on the PDF central values and uncertainties. Second we produced a new PDF set based on the same data and theory input as NNPDF4.0 \footnote{Theory and data has been updated since Ref.~\cite{NNPDF:2021njg}, see appendix A of Ref.~\cite{NNPDF:2024djq} and Ref.~\cite{Cridge:2026jyh} for a description of these changes and their effects.}, but deploying the new methodology, i.e., including the sampling over different trials produced according to the new metric.
By comparing our results with PDFs produced using the old NNPDF4.0 methodology we assess the impact of hyperparameter variations on central values and uncertainties. 

In order to stress test the new framework, we also apply it to a PDF model artificially chosen to be inadequate, in that it misses important physical constraints. We study how the hyperoptimisation algorithm adapts its output accordingly, spotting the necessity to completely change the methodology when a less suitable model is used. 

\subsection{Evolution basis}
\label{sec:evol_basis_res}
\paragraph{Initial distribution of PDF replicas.} 
A PDF fit requires the choice of a specific fitting basis,
i.e., a linearly independent set of flavours to parameterize.
The NNPDF4.0 analysis adopts the so-called evolution basis, defined as
\begin{align}
    \label{eq:evol_basis}
    f_i = \{\Sigma, g, V, V_3, V_8, T_3, T_8, T_{15} \}\,,
\end{align}
with 
\begin{align}
    \label{eq:evol_basis_def}
    &\Sigma = u + \bar{u} + d + \bar{d} + s + \bar{s} + 2c\,, \nonumber\\ 
    & V = \left(u - \bar{u}\right) + \left(d - \bar{d}\right) + \left(s - \bar{s}\right)\,, \nonumber \\
    &V_3 = \left(u - \bar{u}\right) - \left(d - \bar{d}\right)\,, \nonumber\\
    &V_8 = \left(u - \bar{u} + d - \bar{d}\right) - 2 (s - \bar{s})\,, \nonumber\\
    &T_3 =  \left(u + \bar{u}\right) - \left(d + \bar{d}\right)\,, \nonumber\\
    &T_8 = \left(u + \bar{u} + d + \bar{d}\right) - 2 (s + \bar{s})\,, \nonumber \\
    &T_{15} = \left(u + \bar{u} + d + \bar{d} + s + \bar{s}\right) - 3\left(c+\bar{c}\right)\,.
\end{align}
With this choice QCD evolution equations are block diagonal,
with only the gluon and singlet $\Sigma$ PDF mixing under DGLAP, and valence $V_i$ and nonsinglet sea $T_i$ being evolution eigenvectors.

Using the notation introduced in Sec.~\ref{sec:nnpdf_methodology}, the choice of a specific fitting basis determines the flavours $i$ for which probability distributions are introduced. 
By discretizing the problem as per Eq.~\eqref{eq:FunVect}, samples $\mathbf{f}_i$ from the initial PDF distribution are generated according to
\begin{align}
    \label{eq:evol_basis_prior}
    \mathbf{f}_i = A_i \,x^{-\alpha_i}\left(1-x\right)^{\beta_i} \text{NN}_i\left(x; \mathbf{w},\mathbf{\theta}\right)\,,
\end{align}
by drawing random values for the network weights $\mathbf{w}$. Eq.~\eqref{eq:evol_basis_prior} defines the initial probability distribution for the flavours in Eq.~\eqref{eq:evol_basis}, by specifying how to generate samples from it:
$\text{NN}_i$ is the $i$-th output node of a neural network $\text{NN}$, with weights $\mathbf{w}$ and hyperparameters $\mathbf{\theta}$; $A_i$ is a normalization factor ensuring that valence and momentum sum rules are satisfied; $\alpha_i$ and $\beta_i$, denoted as preprocessing exponents, are kept fixed during the fit, and determined iteratively, by varying them in a range which is consistent with theoretical knowledge of the small- and large-$x$ behaviour of the PDF.

The preprocessing exponents play an important role. Considering, for example, the nonsinglet valence distributions $V_i$, their correct small-$x$ behaviour has to be compatible with the condition 
\begin{align}
    \label{eq:V_at_smallx}
    \lim_{x\rightarrow 0} xV_i\left(x\right) = 0\,,
\end{align}
necessary not only to describe the experimental data but also to ensure the correct integrability properties of the distributions.
By choosing $\alpha_{V_i}$ in the range $(0,1)$, this known theoretical feature can be implemented both in the initial replica distribution and during the optimisation: at any point during training, the PDF replicas will show the desired small-$x$ behaviour.
The space of possible solutions which can be accessed during the fit is therefore greatly reduced, making the minimization process much simpler and the final results more accurate in reproducing the known theoretical features.

\paragraph{Hyperparameter space.}
As mentioned at the beginning of Sec.~\ref{sec:hyperscan}, the full methodology is defined by architecture and training hyperparameters:
while the former control specific features of the initial PDF replicas distribution $\text{NN}\left(\mathbf f\right)$, the latter determine the optimisation procedure which brings to the final distribution of trained replicas $p\left(\mathbf f | y, \theta\right)$.
The interplay between these two families of hyperparameters is non-trivial: the specifics of the optimisation procedure will strongly depend on the features of the initial distribution and, additionally, on the theoretical constraints which are built in the neural network and imposed throughout the training. 
Therefore, different architectures coupled with different optimisation procedures could in principle lead to methodologies which are equally good according to the metric of Eq.~\eqref{eq:nnpdf4.1_hyperopt_metric}, leading however to visible effects at the level of the final PDF.
In order to let the hyperoptimisation algorithm explore this possibility, we allow both sets of hyperparameters to vary at the same time. 

The specific settings are reported in Tab.~\ref{tab:hyperparam_space_evol}:
the neural network can present between 2 and 5 layers, with a number of nodes varying between 10 and 25 and activation functions being either sigmoid or hyperbolic tangent; the minimiser used during the training is sampled considering \texttt{Nadam} and \texttt{Adam} optimizers, with variable learning rates and clipnorms; the stopping options are also included in the explored hyperparameter space, allowing to sample among different number of epochs and patience.
In order to facilitate the comparison with the results from NNPDF4.0 we have taken the folds to be the same as in Ref.~\cite{NNPDF:2021njg} (Table.~8).

\begin{table}
\centering
\small
\begin{tabular}{|c|l|c|}
\toprule
\multirow{3}{10em}{\textbf{Architecture}} &  Layers & 2,3,4,5  \\ 
& Number of nodes & min: 10, max: 25   \\ 
& Activation  & sigmoid, tanh\\
\midrule
\multirow{3}{10em}{\textbf{Training}} &  Optimizer & Nadam, Adam  \\ 
& Learning rate & min: $10^{-3}$, max: $10^{-2}$, sampling: uniform   \\ 
& Clipnorm  & min: $10^{-7}$, max: $10^{-5}$, sampling: log\\
\midrule
\multirow{2}{10em}{\textbf{Stopping}} &  Epochs & min: $15 \times 10^{3}$, max: $25\times 10^{3}$  \\ 
& Patience & min: $0.1$, max: $0.2$   \\ 
\bottomrule
\end{tabular}
\caption{\sf Hyperparameter space for hyperoptimisation in the evolution basis.}
\label{tab:hyperparam_space_evol}
\end{table}

\begin{figure}
\begin{center}
\includegraphics[scale=0.45]{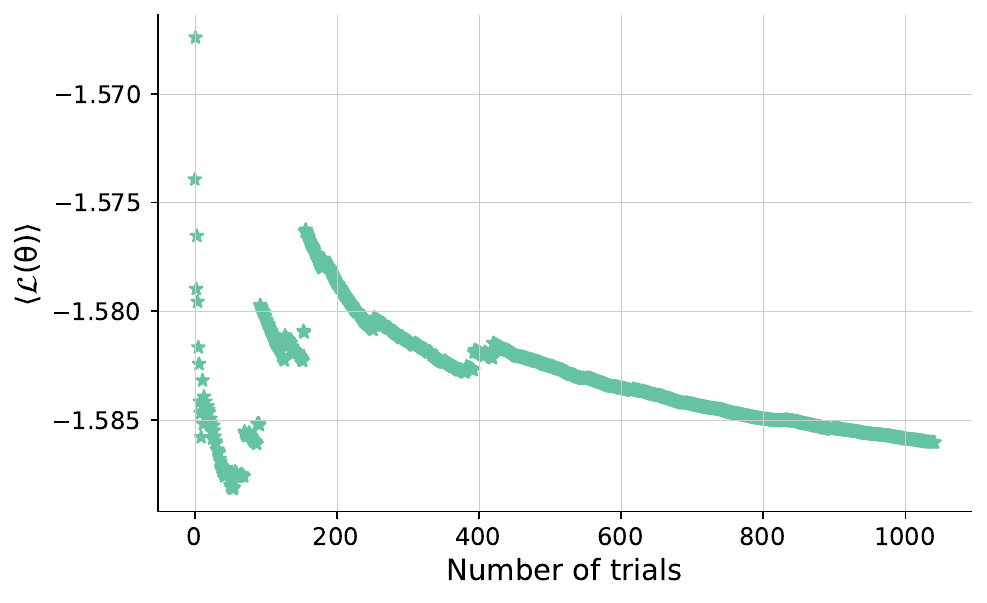}
\includegraphics[scale=0.45]{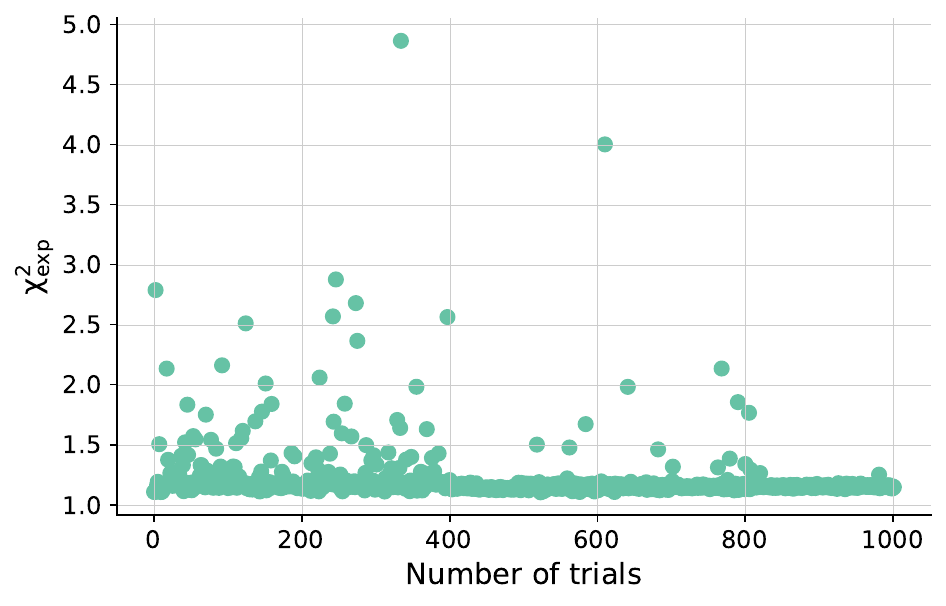}
\end{center}
\caption{\sf Cumulative average of the loss function (left) and $\chi_{\text{exp}}^2$ of the fitted data during $k$-folding (right) as functions of the trials number. The $\chi^{2}$ value includes experimental and PDF uncertainties.}
\label{fig:logp_evol}
\end{figure}

\paragraph{Single trial fits.}
We run the \texttt{hyperopt} library to produce a chain of $\sim 1000$ trials.
The running average of the loss function defined in Eq.~\eqref{eq:nnpdf4.1_hyperopt_metric} is plotted in the left panel of Fig.~\ref{fig:logp_evol}, while in the right panel the experimental $\chi_{\text{exp}}^2$ of the data entering training and validation is reported for each trial, to check on the convergence of the fold fits entering the $k$-folding procedure.
By inspecting such plots we observe how, after an initial thermalisation phase of about $\sim 400$ trials the loss function steadily decreases. The general convergence of the fits to individual folds also becomes more stable after such thermalisation phase, with the number of fits having $\chi^2_{\text{exp}} > 2$ drastically decreasing.

 In Tab.~\ref{tab:best10_evol} we report the 10 best
 trials selected by the hyperoptimisation algorithm\footnote{In this study we consider ensembles of 10 trials, but this is an arbitrary choice for the purposes of this study, in principle one could sample from the entire thermalisation phase according to their figure of merit.}.
For each trial we indicate its position in the chain produced by \texttt{hyperopt} (first column) and the specific hyperparameter settings defined in Tab.~\ref{tab:hyperparam_space_evol}.
In order to assess the dependence of the results on the different hyperparameter settings, we first produce 10 fits, each one based on the methodology of one of the 10 best trials. 
To quantify the relative change in PDF uncertainty we refer to the $\phi$ estimator introduced in Ref.~\cite{NNPDF:2014otw}, defined as 
\begin{align}
    \label{eq:phi}
    \phi = \frac{1}{N_{data}} \sum_{i,j} \left[C^{-1}_y\right]_{ij}\left[C_{PDF}\right]_{ji}\,,
\end{align}
giving the uncertainties and correlations in the theory
predictions using the output PDF, normalized to those of the original data,
averaged over all data. 
In Tab.~\ref{tab:best10_evol_chi2_phi} we report the values of $\chi^2$ as defined in Eq.~\eqref{eq:chi2term}, including the PDF uncertainty, and of $\phi$, for single trial fits based on the 10 best trials of Tab.~\ref{tab:best10_evol}. 
We observe good stability of fit quality and PDF uncertainties across different trials, showing how each of them gives an equivalently good description of the data (given by the value of $\chi^2$) and a similar variance for all ensembles as measured by the $\phi$ estimator.

\begin{table}[tb]
\centering
\small
\begin{tabular}{llllrrr}
\toprule
 Trial  & Architecture & Activation & Optimizer & Learning rate & Clipnorm & Epochs \\
\midrule
925  & [15, 16, 19, 8] & tanh & Adam & 0.00618 & 5.98e-7 & 18704 \\
550  & [13, 16, 18, 8] & tanh & Adam & 0.00816 & 4.67e-7 & 20089 \\
870  & [15, 16, 18, 8] & tanh & Adam & 0.00785 & 6.06e-7 & 19298 \\
760  & [22, 16, 16, 8] & tanh & Adam & 0.00763 & 8.96e-7 & 17993 \\
859  & [15, 17, 17, 8] & tanh & Adam & 0.00850 & 4.97e-7 & 19303 \\
463  & [12, 16, 19, 8] & tanh & Adam & 0.00979 & 1.93e-6 & 18683 \\
729  & [19, 17, 17, 8] & tanh & Adam & 0.00695 & 6.21e-7 & 18880 \\
666  & [12, 12, 17, 8] & tanh & Adam & 0.00491 & 8.02e-7 & 22204 \\
898  & [24, 15, 19, 8] & tanh & Adam & 0.00891 & 4.65e-7 & 19564 \\
438  & [14, 15, 19, 8] & sigmoid & Adam & 0.01000 & 1.95e-6 & 19737 \\
\bottomrule
\end{tabular}
\caption{\sf The 10 best trials selected by the hyperoptimisation algorithm. }
\label{tab:best10_evol}
\end{table}

\begin{figure}
\begin{center}
\includegraphics[scale=0.45]{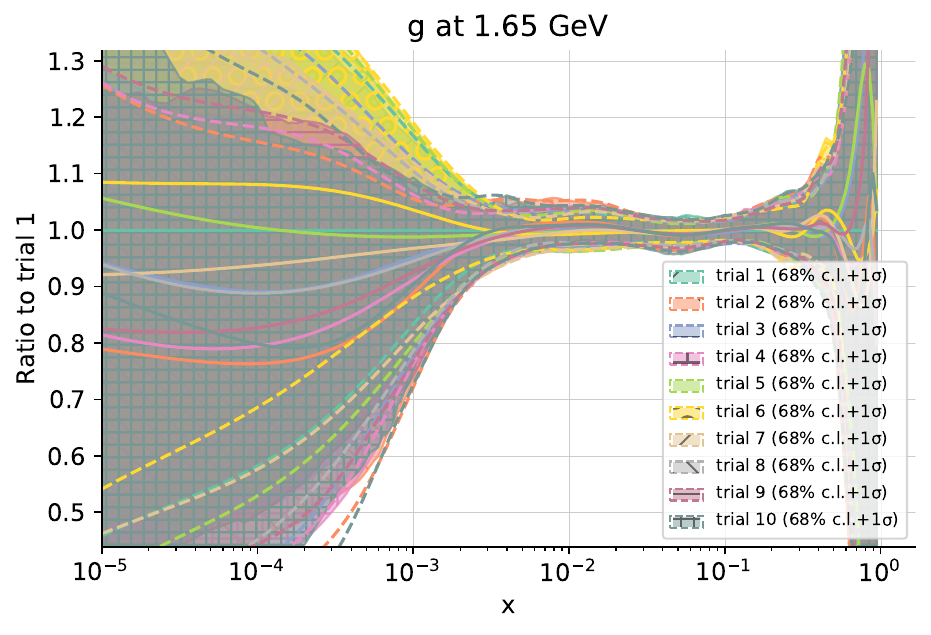}
\includegraphics[scale=0.45]{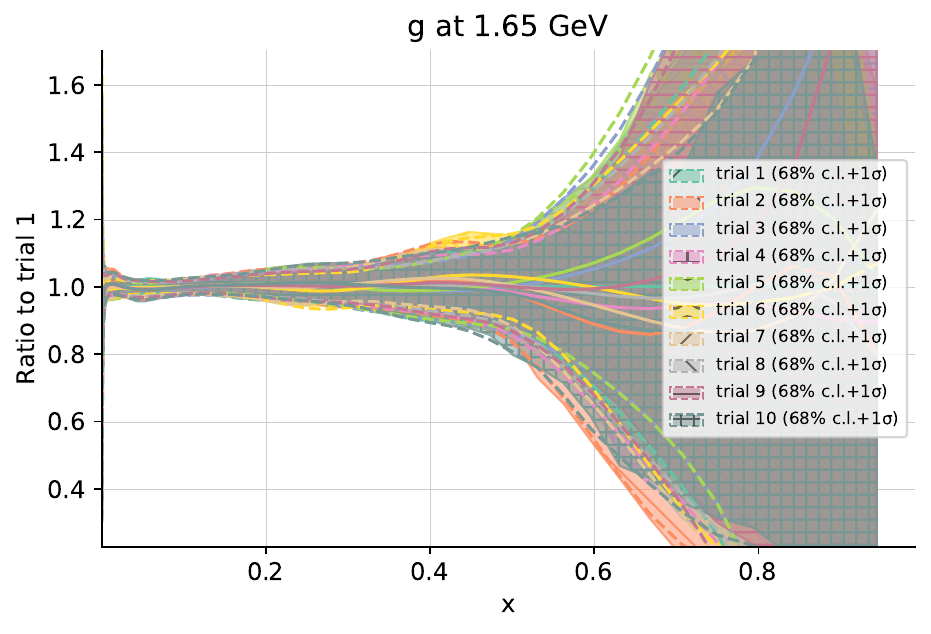}
\end{center}
\caption{\sf Gluon PDF in logarithmic and linear scale from the 10 best trials.}
\label{fig:g_trials_evol}
\end{figure}

In Fig.~\ref{fig:g_trials_evol} we plot the gluon PDFs computed from the 10 best trials, normalized to the central value of trial 1:
a spread of the central values can be observed starting from $x\sim 0.8 \times 10^{-2}$ down to the small-$x$ extrapolation regions, where fluctuations can go up to $\sim 20\%$, and again from $x\sim 0.3$ up to large-$x$ extrapolation, where central value fluctuations can go up to $60\%$.
Regions displaying large central value fluctuations are also characterized by an increased PDF uncertainty, such that different trial results are always compatible within uncertainties.
Nonetheless, given the size of the fluctuations an increase of the total PDF error due to the second term of Eq.~\eqref{eq:mc_error_decomposed} can be expected in such regions upon combination of different trial results.

\begin{figure}
\begin{center}
\includegraphics[scale=0.45]{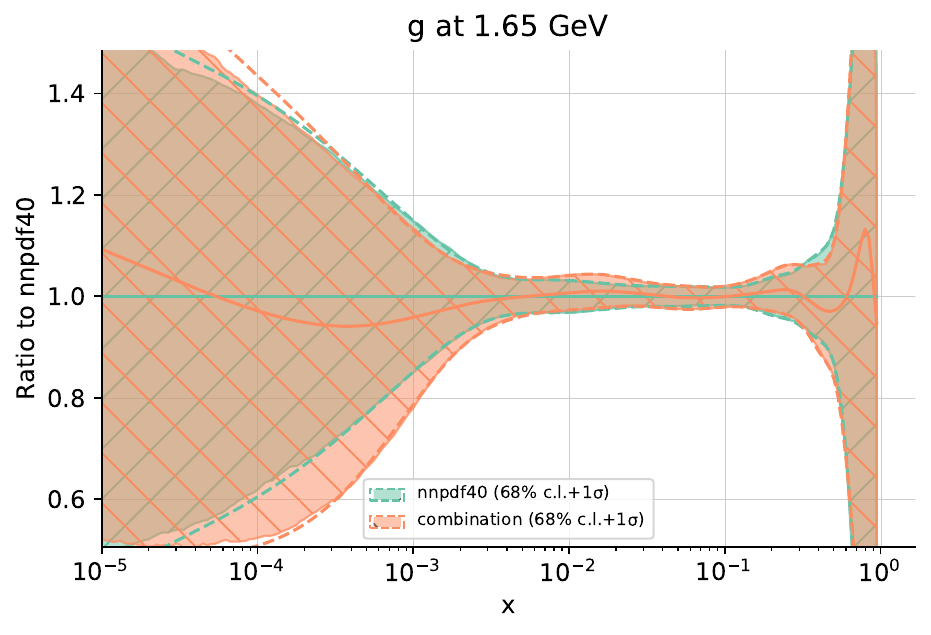}
\includegraphics[scale=0.45]{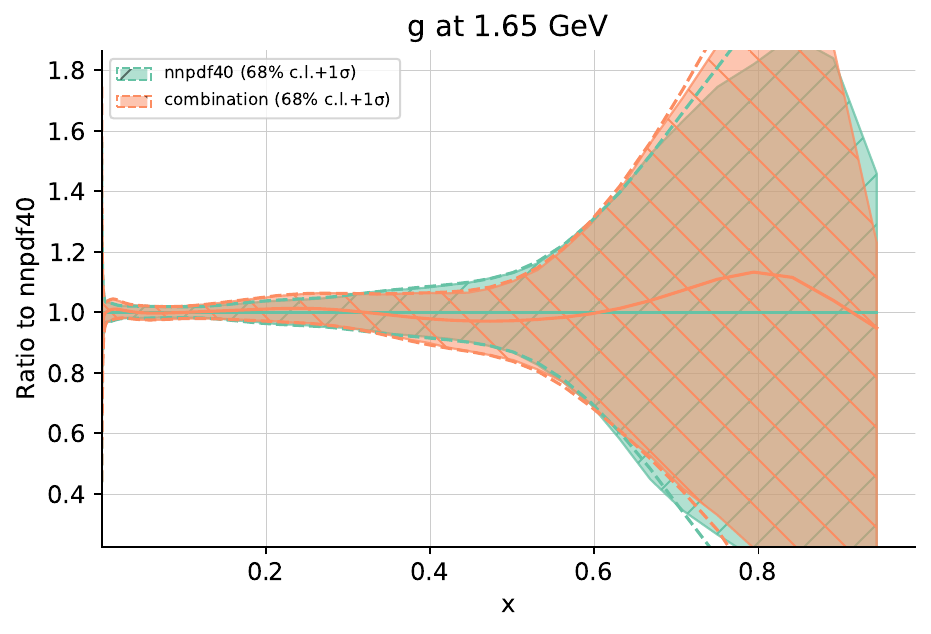}
\includegraphics[scale=0.45]{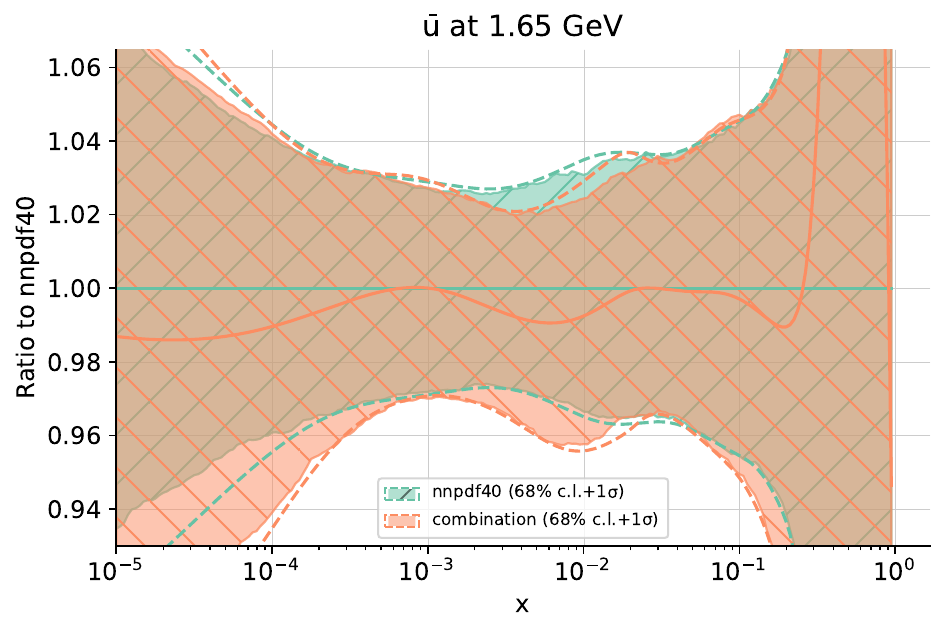}
\includegraphics[scale=0.45]{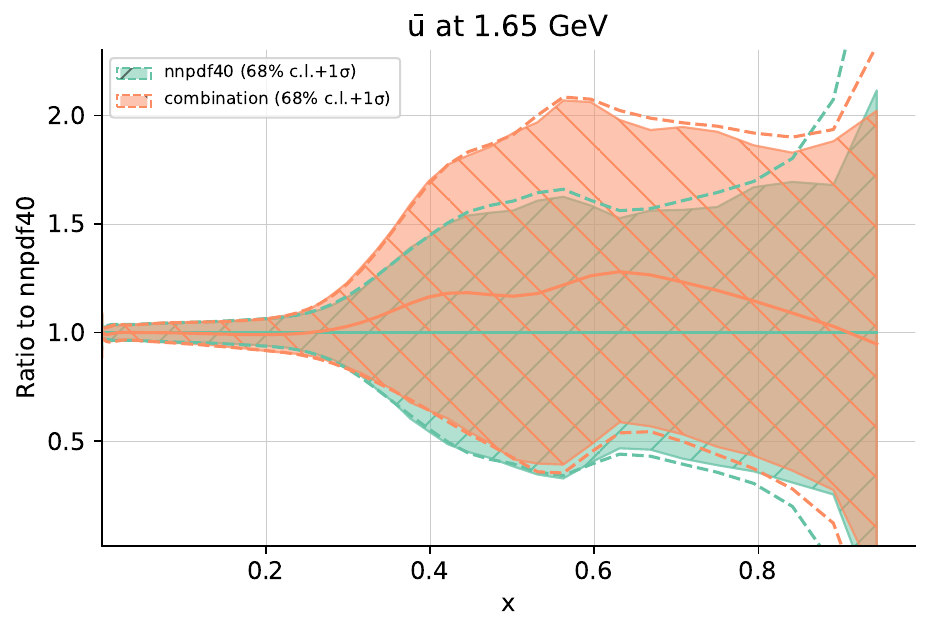}
\end{center}
\caption{\sf Comparison between combination fit and nnpdf40-like fit.}
\label{fig:combination_vs_40_pdfs}
\end{figure}

\paragraph{Combination and comparison with NNPDF4.0.}
Fits produced using the 10 best trials are then combined in a unique fit of 1000 Monte Carlo replicas, which is taken to be the final result.
In Tab.~\ref{tab:combination_vs_40} we report $\chi^2$ and $\phi$ values for the combination and a baseline fit produced using the NNPDF4.0 methodology: we observe an equally good description of the data, and an increase of the PDF error (as given by the $\phi$ parameter) of about 12\%.
In the upper panel of Fig.~\ref{fig:combination_vs_40_pdfs} we report comparison plots between gluon PDFs: the uncertainty shows a marked increase at small-$x$, while remaining about the same in the data and large-$x$ regions.
Looking at individual quarks distributions instead we observe a general increase of the PDF error at both large and small-$x$, as can be observed in the lower panel of Fig.~\ref{fig:combination_vs_40_pdfs}, where the $\bar{u}$ PDF is reported as an example.

\begin{table}
\centering
\small
\begin{tabular}{lrrrrrrrrrr}
\toprule
 & T1 & T2 & T3 & T4 & T5 & T6 & T7 & T8 & T9 & T10 \\
\midrule
$\chi2$ & 1.091 & 1.075 & 1.090 &	1.086 &	1.092 & 1.095 & 1.093	& 1.090 & 1.088 &	1.073 \\
$\phi$ & 0.171 & 0.171 & 0.177 & 0.180 & 0.176 &	0.195 &	0.171 &	0.170 &	0.175 &	0.172 \\
\bottomrule
\end{tabular}
\caption{\sf $\chi^2$ and $\phi$ for the 10 best trials.}
\label{tab:best10_evol_chi2_phi}
\end{table}

\begin{table}
\centering
\small
\begin{tabular}{lccc}
\toprule
 & Combination evolution & Combination flavour & NNPDF4.0 fit  \\
\midrule
$\chi2$ & 1.085  & 1.059 & 1.098  \\
$\phi$ & 0.181 & 0.232 & 0.163  \\
\bottomrule
\end{tabular}
\caption{\sf Values of $\chi^2$ and $\phi$ for the full set of combined trials in both the evolution and flavour basis, compared to a NNPDF4.0 baseline.}
\label{tab:combination_vs_40}
\end{table}

\newpage

\subsection{Flavour basis}
\label{sec:flav_basis}
In this section we repeat the analysis performed in Sec.~\ref{sec:evol_basis_res}, but starting from a different set of independently parameterized flavours, as done across different PDF fitting groups.
We adopt the so-called flavour basis
\begin{align}
    \label{eq:flav_basis}
    \tilde{f}_i = \{g, u, \bar{u}, d, \bar{d}, s, \bar{s}, c \}\,.
\end{align}
Each flavour PDF in this basis is generated using
\begin{align}
    \label{eq:flav_basis_prior}
    \tilde{\mathbf{f}}_i = x^{-1}\left(1-x\right)^{\beta_i} \text{NN}_i\left(x; \mathbf{w},\mathbf{\theta}\right)\,,
\end{align}
with no variable small-$x$ polynomial factor implemented here, unlike Eq.~\eqref{eq:evol_basis_prior}.
Sum rules are imposed by first rotating in the evolution basis~\eqref{eq:evol_basis} and then computing the model normalization, so that samples for the
variables $\mathbf{f}_i$ identifying evolution basis flavours are now given by
\begin{align}
    \label{eq:flav_basis_param_2}
    \mathbf{f}_i = A_i\,\sum_k\text{R}_{ik}\,\,x^{-1}\left(1-x\right)^{\beta_k} \text{NN}_k\left(x; \mathbf{w},\mathbf{\theta}\right)\,,
\end{align}
with $\text{R}_{ik}$ representing a rotation from $\tilde{f}_i$ to $f_i$, and $A_i$ the normalization factors to ensure the validity of sum rules.
Referring to the discussion in Sec.~\ref{sec:evol_basis_res}, it is clear how the main differences with respect to the evolution basis case occur in the small-$x$ region, where the lack of constraint both in the initial distribution of PDF replicas and, most importantly, at any point during training, forces the network to learn this behaviour from the data.
In order to stress test the new hyperoptimisation algorithm, we apply it to the flavour basis case, and we study how results change due to the lack of physical information of this model.


The folds are the same as in the case of the evolution basis studies. Also in this case we allow for variations of both architecture and training hyperparameters. This is crucial for the hyperoptimisation algorithm to work properly by exploring possible interplays between the new architectures and the optimisation, given the change in the initial distribution of PDF replicas and in the space of possible solutions accessible during minimization. 
We have observed that, in order to obtain a good fit to the data, a wider network architecture and a larger number of training epochs are required compared to the evolution basis hyperoptimisation. The sampling hyperparameter space has been therefore adapted accordingly. The specific settings are reported in Tab.~\ref{tab:hyperparam_space_flav}.

\begin{table}
\centering
\small
\begin{tabular}{|c|l|c|}
\toprule
\multirow{3}{10em}{\textbf{Architecture}} &  Layers & 3,4,5  \\ 
& Number of nodes & min: 10, max: 45   \\ 
& Activation  & sigmoid, tanh\\
\midrule
\multirow{3}{10em}{\textbf{Training}} &  Optimizer & Nadam, Adam  \\ 
& Learning rate & min: $10^{-3}$, max: $10^{-1}$, sampling: log   \\ 
& Clipnorm  & min: $10^{-7}$, max: $10^{-4}$, sampling: log\\
\midrule
\multirow{2}{10em}{\textbf{Stopping}} &  Epochs & min: $35 \times 10^{3}$, max: $55\times 10^{3}$  \\ 
& Patience & min: $0.1$, max: $0.2$   \\ 
\bottomrule
\end{tabular}
\caption{\sf Hyperparameter space for hyperoptimisation in the flavour basis.}
\label{tab:hyperparam_space_flav}
\end{table}

\begin{figure}
\begin{center}
\includegraphics[scale=0.45]{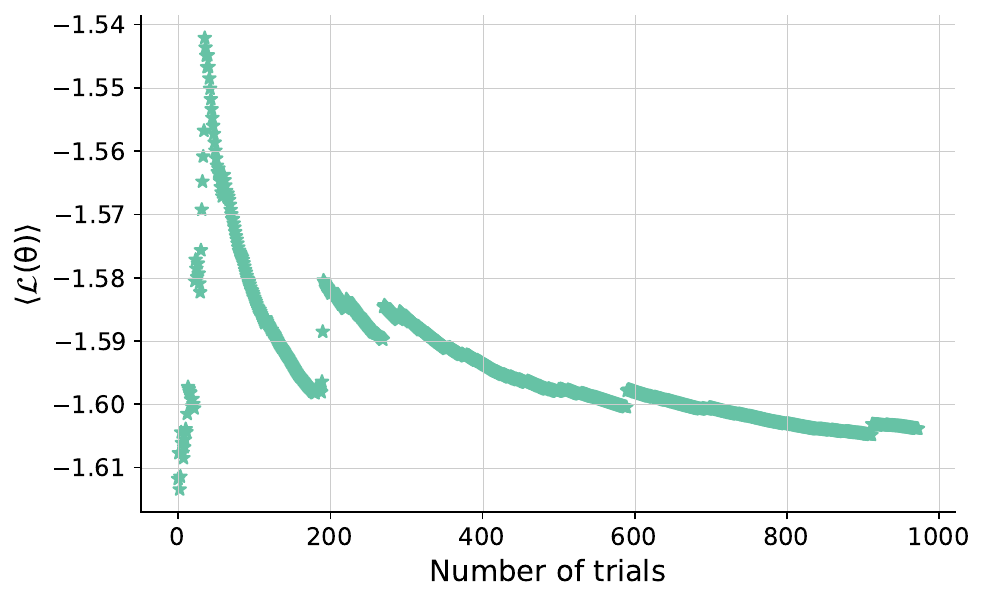}
\includegraphics[scale=0.45]{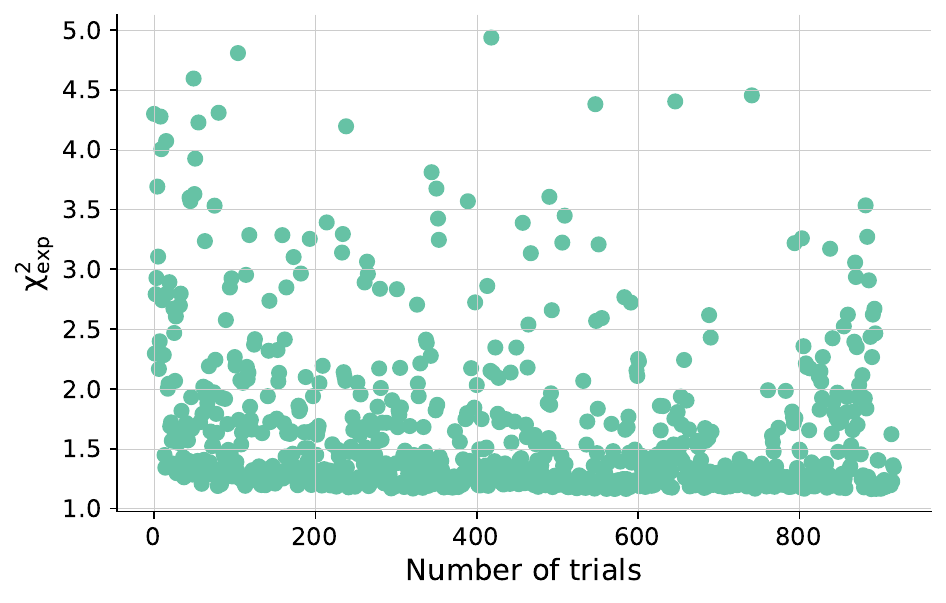}
\end{center}
\caption{\sf Cumulative average of the loss function (left) and $\chi^2$ of the fitted data (right) as functions of the trials number.}
\label{fig:logp_flav}
\end{figure}

\begin{table}
\centering
\small
\begin{tabular}{llllrrr}
\toprule
 Trial  & Architecture & Activation & Optimizer & Learning rate & Clipnorm & Epochs \\
\midrule
710  & [39, 25, 18, 8] & sigmoid & Adam & 0.00234 & 0.00004 & 50727 \\
400  & [43, 36, 21, 8] & sigmoid & Adam & 0.00365 & 0.00010 & 46294 \\
594  & [37, 38, 25, 8] & sigmoid & Adam & 0.00260 & 0.00010 & 50927 \\
682  & [38, 40, 23, 8] & sigmoid & Adam & 0.00287 & 0.00005 & 49113 \\
540  & [38, 24, 19, 8] & sigmoid & Adam & 0.00514 & 0.00008 & 49171 \\
888  & [41, 37, 15, 8] & sigmoid & Nadam & 0.00727 & 0.00002 & 52379 \\
891  & [42, 25, 20, 8] & sigmoid & Adam & 0.00322 & 0.00003 & 46261 \\
413  & [45, 22, 16, 8] & sigmoid & Adam & 0.00367 & 0.00010 & 46411 \\
963  & [40, 39, 15, 8] & sigmoid & Adam & 0.00272 & 0.00007 & 49697 \\
858  & [38, 27, 25, 8] & sigmoid & Adam & 0.00245 & 0.00008 & 48024 \\
\bottomrule
\end{tabular}
\caption{\sf The 10 best trials selected by the hyperoptimisation algorithm in the flavour basis}
\label{tab:best10_flav}
\end{table}

The \texttt{hyperopt} library is again used to produce a chain of $\sim 1000$ trials. In Fig.~\ref{fig:logp_flav} we report the running average of the loss function (left panel) and  $\chi^2_{\text{exp}}$ computed on the data entering training and validation sets (right panel). Comparing the latter with the corresponding evolution basis plot of Fig.~\ref{fig:logp_evol}, we notice how the number of non-converging fits ($\chi^2_{\text{exp}}>2$) still remains very high also when approaching the end of the trials chain, pointing to a general worse convergence of flavour basis trials.
The 10 best trials selected by \texttt{hyperopt} are reported in Tab.~\ref{tab:best10_flav}. These are characterized by a higher number of nodes in the intermediate layers and by a longer training.

\begin{figure}
    \begin{center}
        \includegraphics[scale=0.45]{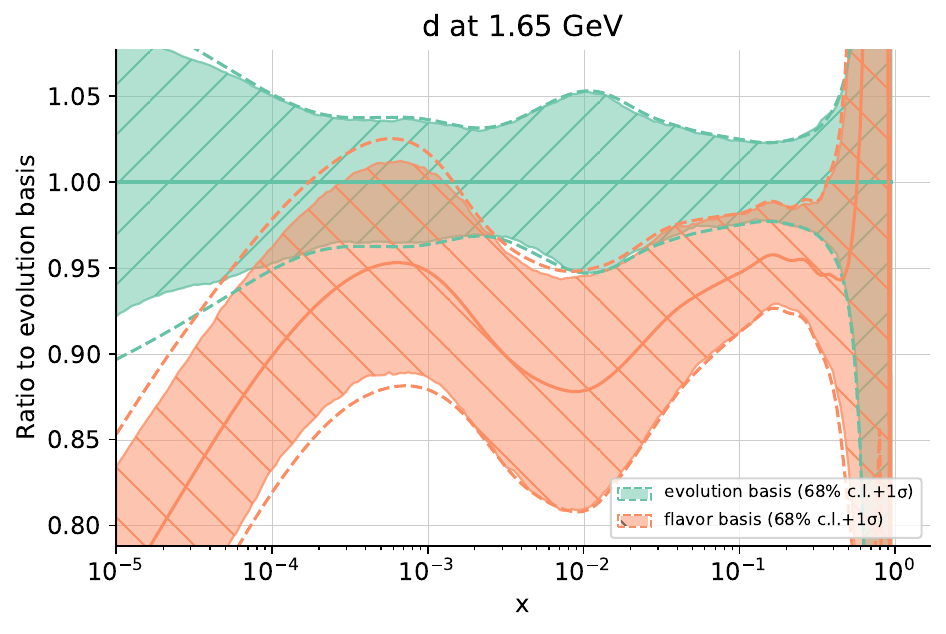}
        \includegraphics[scale=0.45]{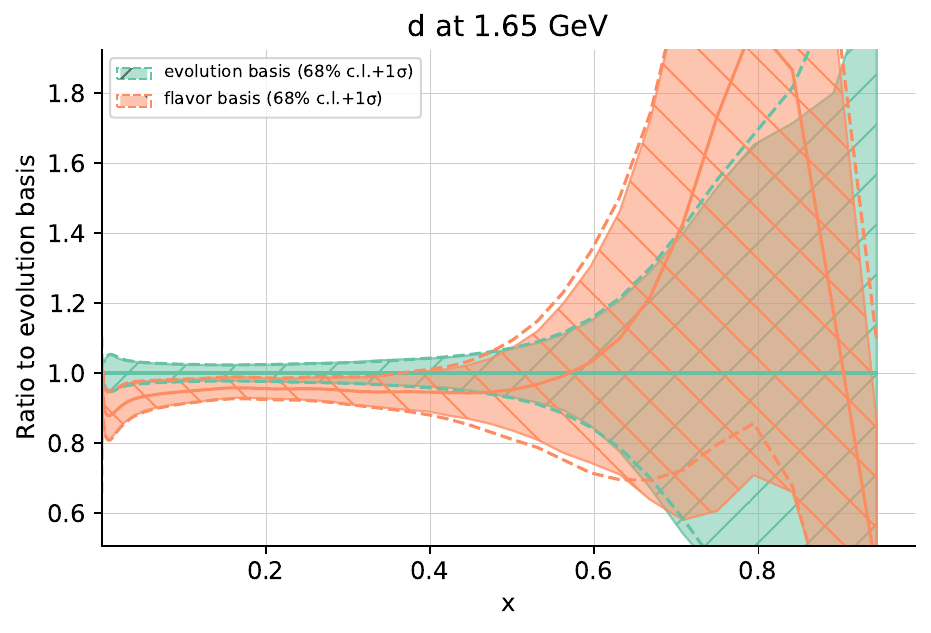}
        \includegraphics[scale=0.45]{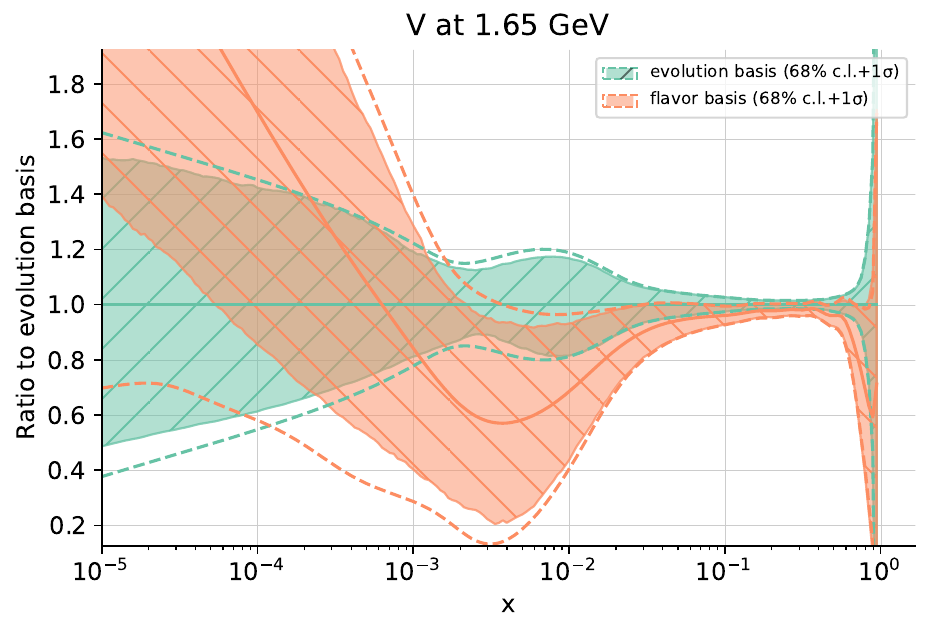}
        \includegraphics[scale=0.45]{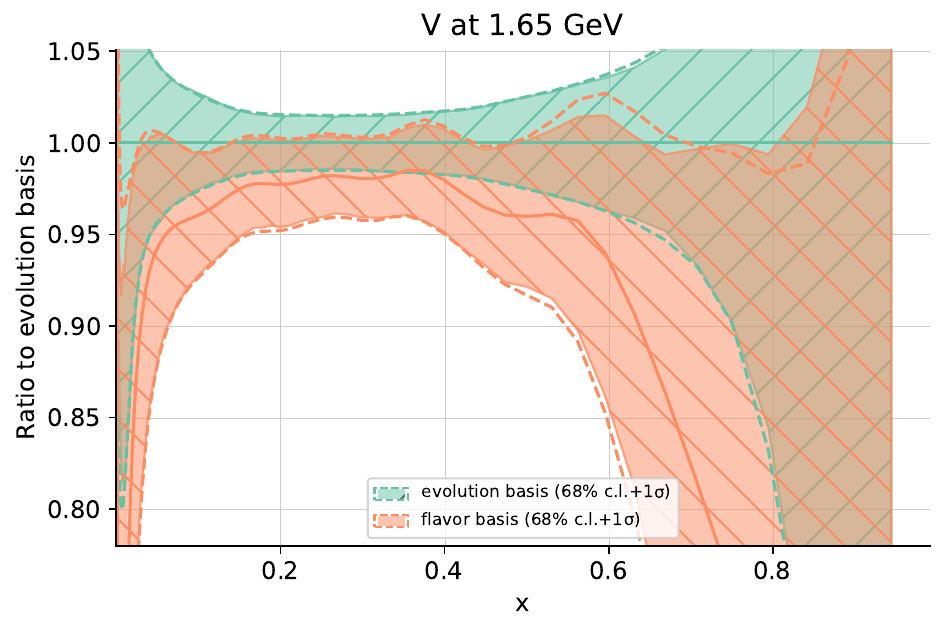}
    \end{center}
\caption{\sf d-quark and V distribution PDFs in both the evolution and flavour basis (PDFs defined by the combination of different trials) at the fitting scale. We observe agreement within uncertainties but each parametrization basis locks into a different behaviour.}
\label{fig:PDF_evol_vs_flav}
\end{figure}

As done for the evolution basis, 10 independent fits are produced using the best trials settings, and a combination of the resulting PDF is produced. The corresponding values of $\chi^2$ and $\phi$ are reported in Tab.~\ref{tab:combination_vs_40}. 
The lower $\chi^2$ value observed for the flavour basis should not be interpreted as a more accurate central PDF, as it is affected by the larger PDF error induced by the additional spread of methodologies,
as it is evidenced by the much larger value of $\phi$ with respect to the evolution basis (or to NNPDF4.0).
Significant differences can be observed also at the PDF level: errors are generally larger in the flavour basis combination, and central values present a different behaviour, with more than $1\sigma$ differences in certain data regions.
As an example, in Fig.~\ref{fig:PDF_evol_vs_flav} we report a comparison between the evolution and flavour basis for the d-quark PDF and Valence distribution, where the differences in their central values at the parametrization scale (1.65 GeV) are particularly evident.
While they have overlapping uncertainties, the overall behaviour in each of the two bases is quite different, with both d-quarks in particular staying at the edge of each other's uncertainties across the whole range in $x$.

\begin{figure}
    \begin{center}
        \includegraphics[scale=0.45]{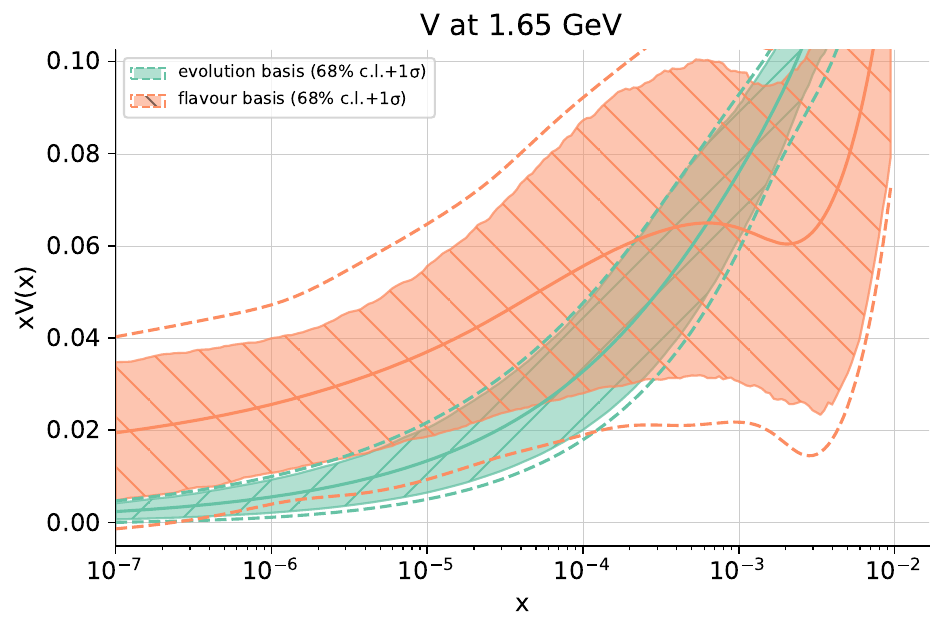}
        \includegraphics[scale=0.45]{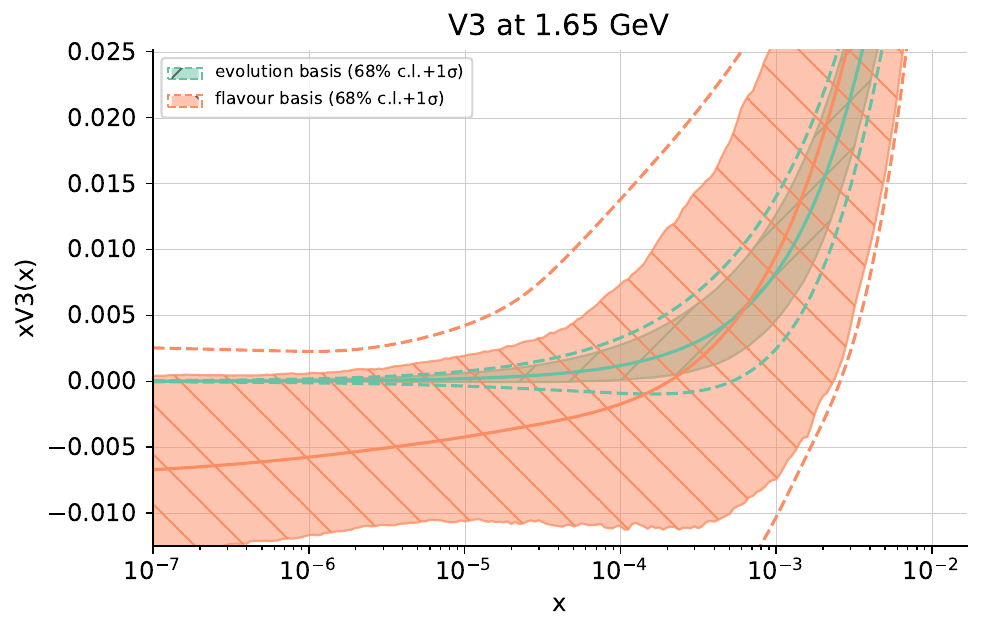}
    \end{center}
\caption{\sf PDFs for both the $V$ and $V_3$ distributions in both the evolution and flavour basis. As it can be readily be observed in this figure, while in the evolution basis the valence-like distributions go to 0 at small $x$ by construction, in the flavour basis this constraint is absent and the network is unable to learn it without the necessary data.}
\label{fig:PDF_evol_vs_flav_zoom_in}
\end{figure}

As mentioned above, the flavour-basis network must learn from the data alone certain theoretical constraints, in a region in which data are scarce, while
when working in the evolution basis this information is incorporated into the model, and the network starts from a physically motivated prior.
Let's consider, for example, the property expressed in Eq.~\eqref{eq:V_at_smallx}.
This must arise from a precise cancellation between different quark and antiquark PDFs at small-$x$, as can be seen by the definition of valence distributions in Eq.~\eqref{eq:evol_basis_def}.
When working in the flavour basis, such cancellation has to be learnt by the network directly from the experimental information, in a region where data provide limited constraining power, which is a very difficult task for any numerical optimization.
This is seen explicitly in Fig.~\ref{fig:PDF_evol_vs_flav_zoom_in} in which the $V$ and $V_3$ distributions are shown, with a focus on the small-$x$ region:
while some replicas in the flavour basis do produce a physical result and are close to zero, the central value is still non-zero at values of $x\sim 10^{-7}$.

It is then remarkable how the hyperoptimisation algorithm manages to select a methodology for the flavour basis model which is still able to get sensible results, despite the lack of any small-$x$ constraints, by spotting the necessity of having a more complex network coupled with a much longer training.
The latter is required to explore a wider space of possible solutions, which now includes unphysical configurations in which the small-$x$ cancellations fail, which were ruled out in the case of evolution basis fits by construction.
When it comes to the accuracy of the final results however, it is clear how evolution basis fits should be preferable, since known theoretical constraints are built in the model itself.

\begin{figure}
    \begin{center}
        \includegraphics[scale=0.45]{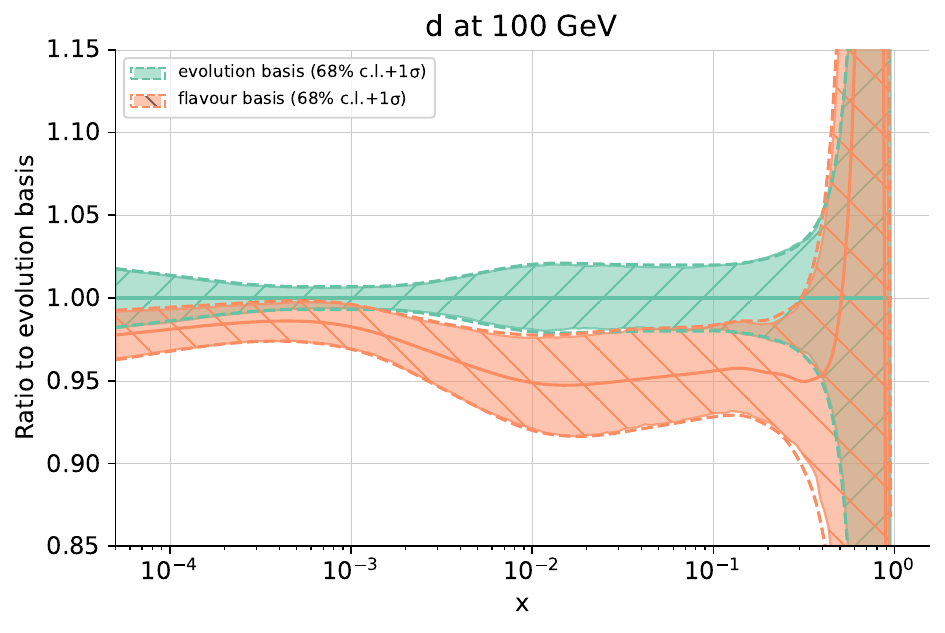 }
        \includegraphics[scale=0.45]{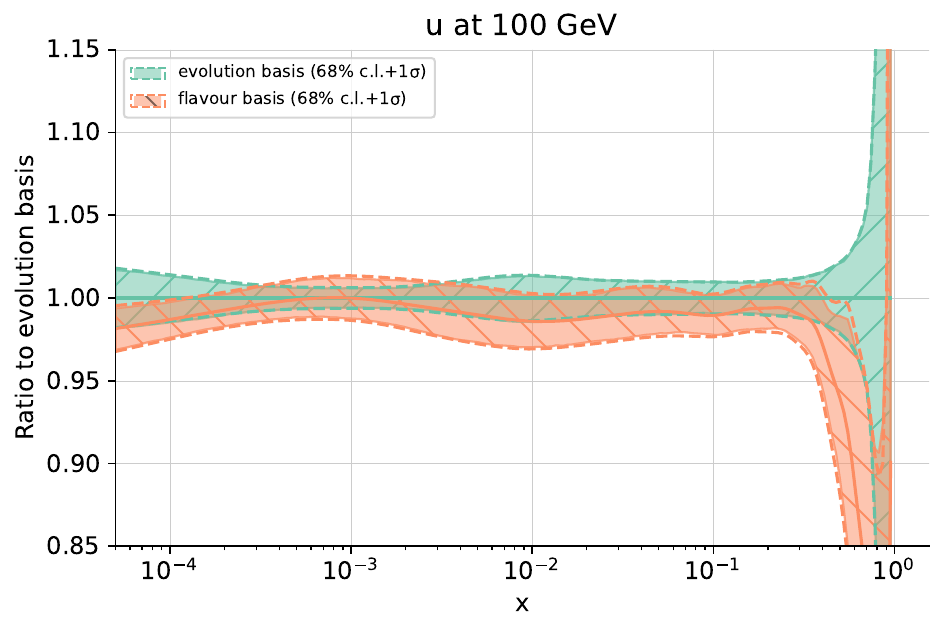 }
    \end{center}
\caption{\sf d- and u-quarks PDFs in both the evolution and flavour after evolution up to 100 GeV, to be compared with Fig. 8.6 of Ref.~\cite{NNPDF:2021njg}. We confirm Ref.~\cite{NNPDF:2021njg} results, with slightly better compatibility due to the enlarged uncertainties.}
\label{fig:PDF_evol_vs_flav_evolved_nnpdf40}
\end{figure}

The tensions observed between evolution and flavour basis results of Sec.~\ref{sec:evol_basis_res},~\ref{sec:flav_basis} are qualitatively similar
to those observed in NNPDF4.0~\cite{NNPDF:2021njg} when the evolution and flavour basis were compared, take as an example Fig.~\ref{fig:PDF_evol_vs_flav_evolved_nnpdf40} in which the d- and u-quarks are shown at $Q=100$ GeV as it is done in Fig. 8.6 of Ref.~\cite{NNPDF:2021njg}.
In a recent study from Ref.~\cite{Cruz-Martinez:2026rct}, in which the MSHT and NNPDF methodologies were compared, similar differences can be observed\footnote{We note however how the MSHT basis is likely not to be affected by the same limitations as the flavour basis model, since valence distributions are parameterized directly.}. 

In summary, also for the flavour basis we confirm the results already seen in Ref.~\cite{NNPDF:2021njg}, obtaining a similar behaviour with slightly larger uncertainties.
We also conclude that, while the hyperoptimisation procedure is very powerful, it still cannot overcome physical limitations of a given model without considerable efforts:
the bigger uncertainties displayed by flavour basis PDF should not be interpreted as the consequence of a more conservative model, but rather as the consequence of a less accurate one, which misses important physical constraints and relies on data and fitting methodology to partially reconstruct them.
A more focused study, outside of the scope of this article, would be necessary to truly put the flavour basis at the same level of accuracy as the evolution basis.
Some possibilities are the use of a penalty term, as it is already done in NNPDF for integrability, to enforce the desired behaviour of the valence distribution, or a flavour-like basis in which the right behaviour can be enforced by construction.



\section{Conclusions}
\label{sec:conclusions}
In this paper we have introduced the new hyperoptimisation algorithm which will be used in the context of future PDF releases by the NNPDF collaboration. Starting from the studies already performed in Refs.~\cite{NNPDF:2021njg, Cruz-Martinez:2024wiu} we have introduced a new hyperoptimisation metric, alongside a new method to sample from the space of hyperparameters. Our work allows to take into account the effects of methodological choices in the final PDF central values and uncertainty, by implementing a $k$-folding algorithm based on the expression of the full likelihood, a task made possible by the more general access to GPU computing.

The new metric and hyperparameters sampling deliver results compatible with previous NNPDF releases, giving however mildly larger PDF uncertainties, quantified in Sec~\ref{sec:evol_basis_res} by using PDF plots and the $\phi$ estimator.
We conclude that the new hyperoptimisation algorithm largely confirms results based on the old NNPDF4.0 methodology, despite delivering a PDF with somewhat larger uncertainties.

In the final section of the paper we have stressed test the algorithm by applying it to a different, less accurate model.
By comparing the results produced using the new hyperoptimised methodology in the context of evolution and flavour basis fits, we have shown how the algorithm successfully captures the different pattern between architecture and optimisation in the two bases, spotting the need for a more complex structure when physical constraints are not enforced.
Nevertheless, the discrepancies between evolution and flavour basis results show how the new hyperoptimisation algorithm cannot fully reconcile tensions between models having different accuracy.
Our results both confirm and further motivate the choice of using the evolution basis for future PDF releases over 
the flavour basis, as the evolution basis provides a more natural framework to incorporate physical information into 
the model.

\section*{Acknowledgments}
We thank Stefano Forte for a careful critical reading of the manuscript, Luigi Del Debbio, Amedeo Chiefa and Juan Rojo for discussions and comments, and the members of the NNPDF collaboration for useful inputs.
The authors acknowledge the use of computing resources on the Leonardo supercomputing cluster at CINECA, Italy, which were instrumental for the large-scale GPU-based hyperparameter optimisation performed in this work. These resources were provided under the INFN project PML4HEP.
This work is also part of the activities of the COST Action CA24146 “Machine Learning and Quantum Computing for Future Colliders” (MLQC4FC).
%
J.C-M. acknowledges support from the Ramón y Cajal program grant RYC2023-043794-I funded by MCIN/AEI/10.13039/501100011033 and by ESF+. 
TG acknowledges support from the European Union under the MSCA fellowship (Grant Agreement N. 101149419) Bayesian tools for one-dimensional and three-dimensional hadron structure (Bayhadron).
This project has also received funding from the French Agence Nationale de la Recherche (ANR) via the grant
ANR-20-CE31-0015 (“PrecisOnium”) and was also partly supported by the French CNRS via the COPIN-IN2P3 bilateral 
agreement and via the IN2P3 project “QCDFactorisation@NLO”.


\appendix
\section{TPE minimisation algorithm: a Bayesian optimisation framework }
\label{app:TPE}

As mentioned in Sec.~\ref{sec:sampling}, TPE is a Bayesian optimisation algorithm constructing a probabilistic surrogate model for the objective function to be minimized in Eq.~\eqref{eq:nnpdf4.1_hyperopt_metric}, and using it to  select the most promising next evaluation point. TPE models the inverse density $p(\theta|\mathcal{L})$ (as opposed to $p(\mathcal{L} | \theta)$ in standard approaches), which is better suited for complex and conditional search spaces.
At a given iteration, the algorithm has accumulated a set of previous evaluations $\{ (\theta_j, \mathcal{L}_j) \}_{j=1}^n$. A threshold $\mathcal{L}^\ast$ is defined as a quantile (typically the 10--20\% best observations) of the observed losses. The hyperparameter density is then decomposed as
\begin{equation}
p(\theta | \mathcal{L}) =
\begin{cases}
\ell(\theta) & \text{if } \mathcal{L} < \mathcal{L}^\ast, \\
g(\theta) & \text{if } \mathcal{L} \ge \mathcal{L}^\ast ,
\end{cases}
\end{equation}
where $\ell(\theta)$ and $g(\theta)$ are the densities of ``good'' and ``bad'' configurations, respectively. 
Both densities are modeled non-parametrically using Parzen estimators, i.e.\ kernel density estimators 
constructed from the sampled hyperparameters in the two subsets.

This reformulation allows one to express the Expected Improvement (EI) acquisition function in a particularly 
simple form. Up to a normalization constant, maximizing EI is equivalent to maximizing the ratio
\begin{equation}
\frac{\ell(\theta)}{g(\theta)} .
\end{equation}
New hyperparameter configurations are therefore sampled preferentially in regions where the probability of belonging to the good set is large compared to the bad set. In practice, candidate points are drawn from $\ell(\theta)$, and the one maximizing $\ell(\theta)/g(\theta)$ is selected for evaluation.

In our setup, each TPE iteration proceeds roughly in four steps. A candidate hyperparameter configuration
$\theta$ is proposed by maximizing the EI criterion derived from $\ell(\theta)$ and $g(\theta)$.
For this configuration, the neural networks are trained independently on each fold, and the fold likelihoods 
of Eq.~\ref{eq:log_likelihood} are computed. The average loss $\mathcal{L}(\theta)$ is evaluated and appended 
to the history. And finally, the densities $\ell(\theta)$ and $g(\theta)$ are updated with the new observations.

\bibliography{references}{}
\bibliographystyle{JHEP}
\end{document}